\documentclass[a4paper,11pt]{article}
\usepackage{pos}

\title{Flow and correlations measurements in
small and large systems}

\author*[a,b]{Lucia Anna Tarasovi\v{c}ová}
\onbehalf{for the ALICE, CMS and ATLAS Collaboration}

\affiliation[a]{Pavol Jozef Šafárik University,\\
Šrobárova 2, 041 54 Košice, Slovakia}
\affiliation[b]{Universität Münster,\\
Wilhelm-Klemm-Straße 9, Münster, Germany}

\emailAdd{lucia.anna.husova@cern.ch}

\abstract{Measurements of flow coefficients and correlations between different types of particles are used to characterise the~properties of the~quark--gluon plasma created in heavy-ion collisions. Moreover, these precise measurements became a key observable in understanding the~possible origin of the~collective-like behaviour in small collision systems.
Recent results of flow and correlations measurements of light and heavy hadrons, in pp, p--Pb, and Pb--Pb collisions are presented. }

\FullConference{The Eleventh Annual Conference on Large Hadron Collider Physics (LHCP2023)\\
 22-26 May 2023\\
 Belgrade, Serbia\\}

\begin{document}
\maketitle

\section{Introduction}

Under the~extreme conditions created in heavy-ion collisions, like high temperature and pressure, the~quark and gluon degrees of freedom manifest in a~deconfined state of matter, quark--gluon plasma (QGP). 
Multiple effects of the~QGP, like collective expansion, strangeness enhancement, jet quenching, or hadron energy loss~\cite{alice}, are measured at the LHC.
These provide quantitative description of the~QGP properties.
Measurements in smaller collision systems, like p--Pb and pp collisions, reveal similarities with heavy--ion collisions. 
Moreover,  a~continuous evolution of some observables, which are interpreted as the~QGP manifestation in the final state, with multiplicity from small to large collision systems is observed. 
This points to a possible creation of a~small QGP-like system also in small collision systems with large final state multiplicity. 

The measurement of the~flow coefficients and correlations between different types of particles can help not only to better constrain the~QGP properties in heavy-ion collisions, such as the~viscosity or the~hadronisation time of different particle species, but also to investigate the~origin of the~collective-like behaviour and its limits in the~small collision systems.

\section{Results}

Balance functions (BFs), which measure the~probability that a~charged particle produced in the~collision will be accompanied by another particle of opposite charge somewhere in the~phase space, serve as a~tool for studies of particle production mechanisms, transport of balancing charges, and production time~\cite{BF_alice}. 
The~BFs of unidentified charged particles and pions, measured by the~CMS and ALICE Collaborations, respectively, show a similar narrowing for the~near-side peak from peripheral to central Pb--Pb collisions. 
This observation is consistent with radial flow effects and the~two-stage quark production scenario~\cite{BF_alice,BF_cms}. 
Qualitatively similar narrowing was observed also in p--Pb collisions, suggesting the~presence of the~radial flow in small collision systems~\cite{BF_cms}.

Many different measurements of anisotropic flow coefficients and their correlations have showed that QGP behaves as an~almost perfect fluid with small shear viscosity per one unit of entropy density, $\eta/s$. 
A~specific type of transverse momentum differential two-particle correlation function, $G_2$ (defined in Ref.~\cite{g2}), has been shown to be sensitive to $\eta/s$. 
The~ALICE Collaboration performed a systematic study of the azimuthal and longitudinal width of both, charge dependent and charge independent $G_2$, as a function of the~charged-particle multiplicity in pp, p--Pb, and Pb--Pb collisions. A~continuous azimuthal narrowing with multiplicity of both correlators is observed, which is consistent with $\langle p_{\mathrm{T}}\rangle$ increase with multiplicity in all collision systems.  
In Pb--Pb collisions, an~around 24\% increase of the~longitudinal width of the charge independent correlator with multiplicity is observed, consistent with viscous effects of long-lived QGP with small $\eta/s$. On the~other hand, it undergoes a~slight narrowing in pp and a~slight broadening in p--Pb collisions in the~longitudinal direction, suggesting that these systems may not live long enough for viscous forces to cause the broadening, even in a~case that QGP would be formed~\cite{g2}. Another explanations are explored by the~comparisons with various models without QGP-like medium, but none of the~considered models is able to give a proper description of the~azimuthal and longitudinal widths of both correlators.

\begin{figure}[b!]
    \centering
    \includegraphics[width=0.86\textwidth]{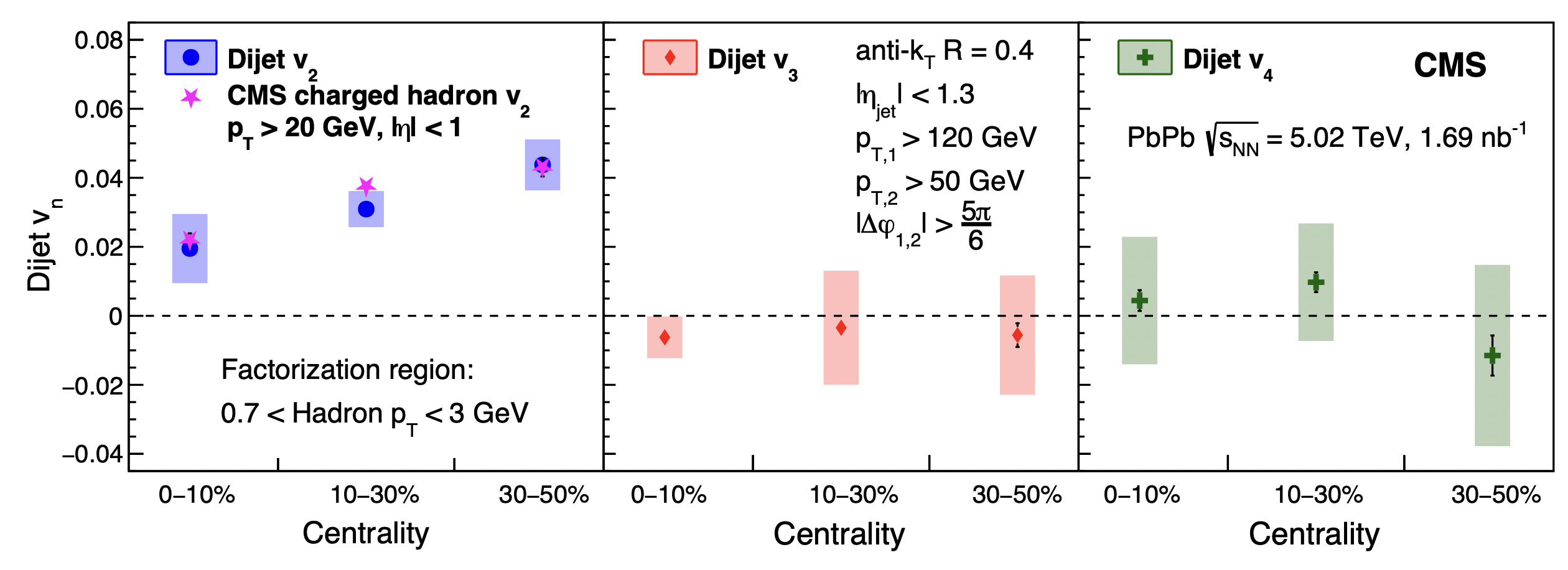}
    \caption{Dijet $v_2$ (left), $v_3$ (middle), and $v_4$ (right) as a function of collision centrality measured in Pb--Pb collisions at $\sqrt{s_{\mathrm{NN}}}$~=~5.02~TeV~\cite{flow_cms}.}
    \label{fig:flow_cms}
\end{figure}

In heavy-ion collisions, the initial spatial anisotropies are translated via interactions during the system evolution into the anisotropies in the final particle distributions in the~momentum space. 
These anisotropies are measured by the~coefficients in the~Fourier expansion of the~distribution of the~azimuthal angle of particles with respect to the~symmetry plane.
The~second coefficient, $v_2$, is called elliptic flow and is caused by the~increased pressure in the in-plane direction due to the~almond shape of the~overlap region of the~two colliding nuclei. 
The~coefficients of higher order are rather generated by the~fluctuations of the~initial state geometry. 
The~medium expansion is affecting mostly the~produced soft particles. 
Nevertheless, the~ATLAS Collaboration observed an~increase of the~elliptic flow coefficient with decreasing centrality from central to semicentral collisions for particles with $p_{\mathrm{T}}$ up to  200 GeV/$c$, while $v_3$ and $v_4$ of these particles is compatible with zero~\cite{flow_atlas}. 
This can be explained by the~different path length connected to the~overall elliptic geometry, which the~high momentum particles need to pass through the~QGP.
While the~distance in the~in-plane direction is shorter, the~path is longer in the~out-of-plane direction. 
Thus, the~hard particles can interact longer with the~medium and loose more energy leading to the~observed anisotropy. 
This explanation is supported also by the~measurement of flow coefficients of dijets in Pb--Pb collisions performed by the~CMS Collaboration, shown in~Fig.~\ref{fig:flow_cms}.

\begin{figure}
    \centering
    \includegraphics[width=0.44\textwidth]{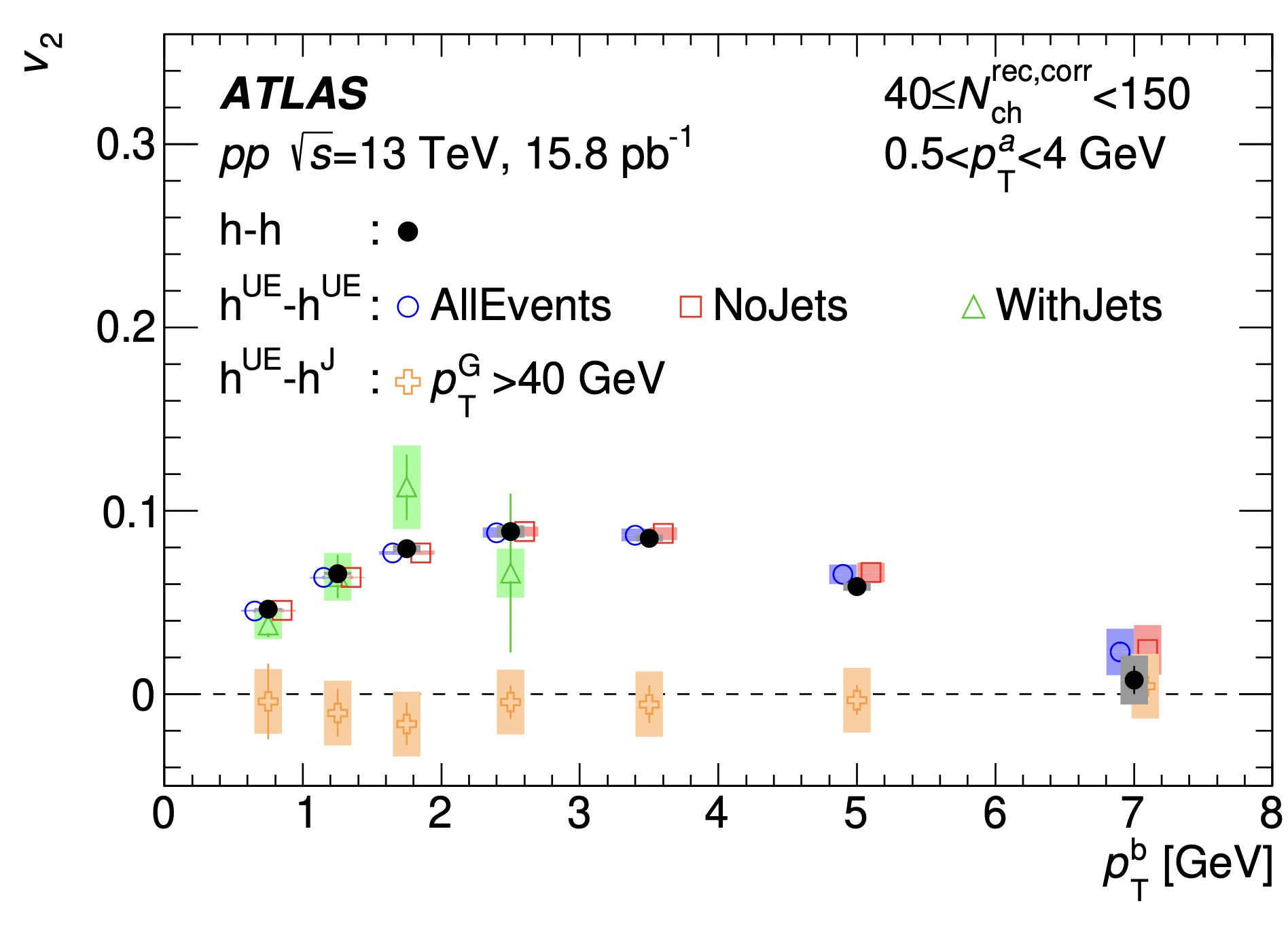}
    \includegraphics[width=0.44\textwidth]{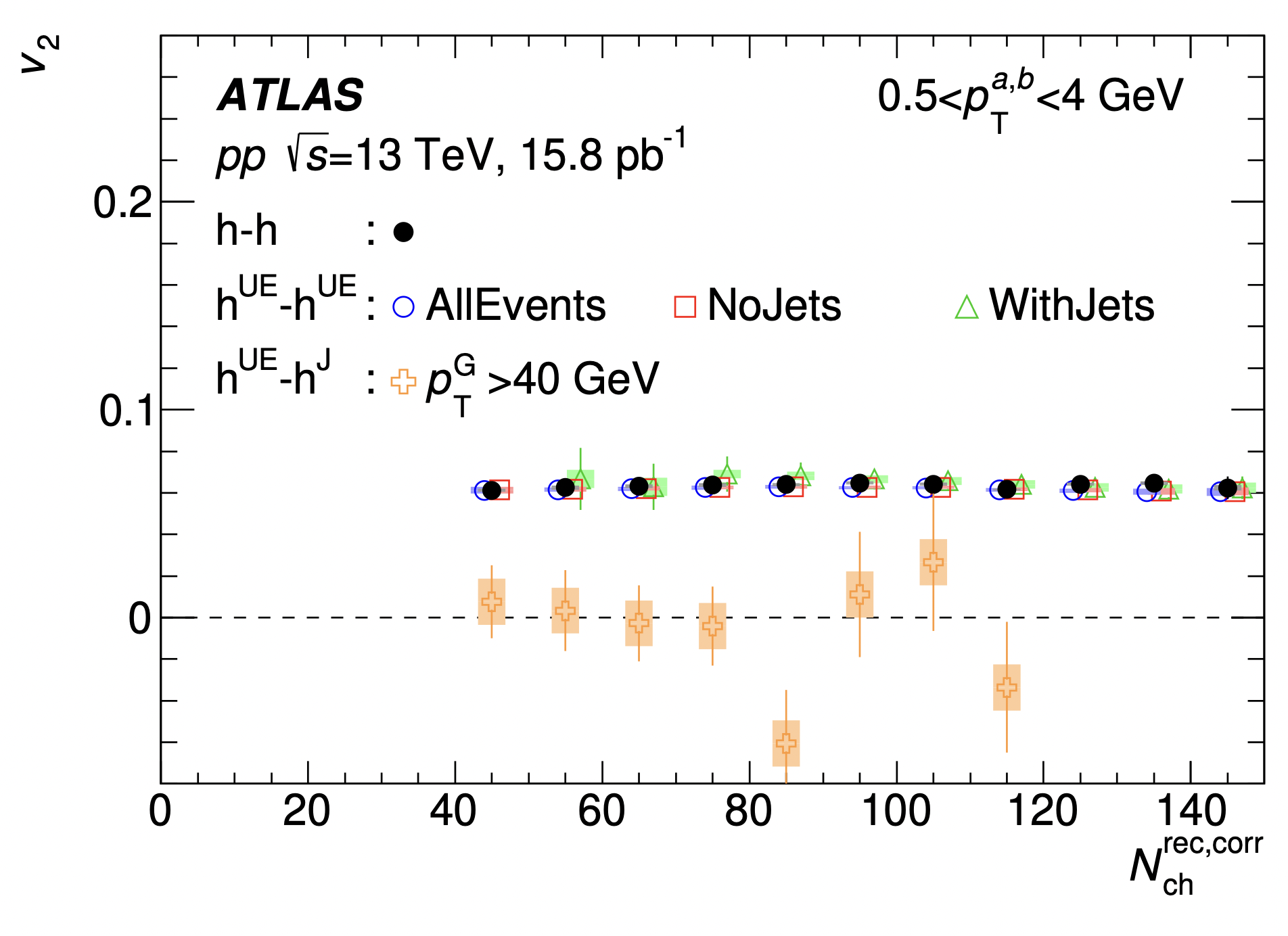}
    \caption{Elliptic flow coefficient measured in pp collisions for jet and underlying event particles as a function of $p_{\mathrm{T}}$ (left) and event multiplicity (right)~\cite{flow_atlas_pp_jets}.}
    \label{fig:flow_atlas_pp}
\end{figure}

Non-zero flow coefficients were observed in many measurements in small systems~\cite{ALICE_flow_allSys,ATLAS_flow_smallSyst}. 
The~origin is still not clear and in order to investigate the~hard fragmentation contribution to $v_2$ in pp collisions, the~ATLAS Collaboration performed a measurement of the~elliptic flow coefficient for inclusive, jet, and out-of jet particles~\cite{flow_atlas_pp_jets}.
The~results are summarised in Fig.~\ref{fig:flow_atlas_pp}. 
It can be observed that while the~$v_2$ of jet particles is compatible with zero, the~elliptic flow coefficient for inclusive and out-of-jet particles with $p_{\mathrm{T}}$ up to 7 GeV/$c$ is non-zero through a~wide range of multiplicities. 
Moreover, a~presence of a~high-$p_{\mathrm{T}}$ jet in a~collision does not affect the~magnitude of $v_2$.
These observations suggests that the~$v_2$ in small collision systems is dominated by soft particles and no indication of (path-length) energy loss of hard particles is present in contrary to Pb--Pb collisions. 

\begin{figure}[b!]
    \centering
    \includegraphics[width=0.55\textwidth]{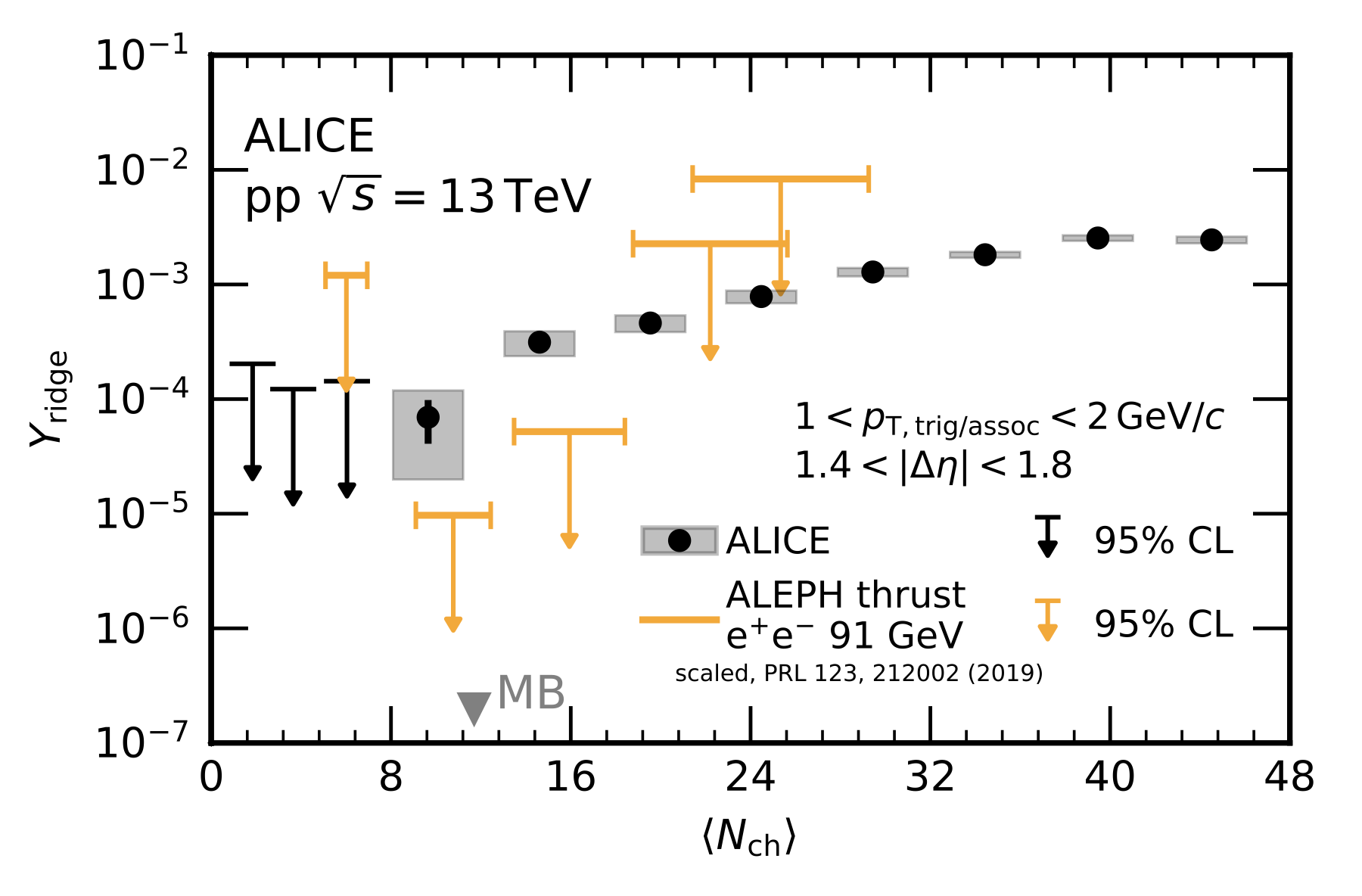}
    \caption{Near-side ridge yield measured in low multiplicity pp collisions at $\sqrt{s}$~=~13~TeV compared with ALEPH $e^+e^-$  results~\cite{ALICE_ridge}.}
    \label{fig:ridge}
\end{figure}

Previous measurement covers rather pp collisions with high final state multiplicities. In order to identify the~minimal size of a collision still showing signs of collectivity, the~ALICE Collaboration performed a~measurement of the~near-side ridge yield visible in long range correlation functions, which is manifestation of the~collective flow. The~measurement shows non-zero ridge yield down to very low multiplicities. 
Moreover, from the comparison with the~same observable measured in $e^+e^-$ annihilations at similar multiplicities (Fig.~\ref{fig:ridge}), it can be concluded that additional processes must occur in hadronic collisions, as the~ridge yield is substantially larger in pp collisions. 

\section{Conclusion}

The~correlation and flow measurements continue to bring insights into the~understanding of both, small and large, collision systems. In Pb--Pb collisions, the~$G_2$ measurement confirms a~creation of a~deconfined matter with small $\eta/s$ while the~BFs support the~two-stage quark production scenario with a~radial expansion in between. Flow coefficient measurements of jets and jet particles show that their~elliptic flow coefficient is induced by the~path length dependent energy loss while they are not sensitive on the~higher order fluctuations. 
A~collective expansion scenario in small systems is supported by the~narrowing of the~peak width of BF and $G_2$ correlation functions of low $p_{\mathrm{T}}$ hadrons, non-vanishing $v_2$ which becomes compatible with zero for jet particles, and a~significantly higher long-range ridge yield than the~one measured in $e^+e^-$ collisions at the~same multiplicity. Nevertheless, viscous forces seem not to have enough time to develop and affect the~$G_2^{CI}$ longitudinal width in the~same manner as in Pb--Pb collisions.


\begin{thebibliography}{99}
\bibitem{alice} ALICE Collaboration, arXiv:2211.04384
\bibitem{BF_alice} ALICE Collaboration, Phys.Lett.B 833 (2022) 137338
\bibitem{BF_cms} CMS Collaboration, arXiv:2307.11185
\bibitem{g2} ALICE Collaboration, Phys.Rev.C 107 (2023) 5, 054617
\bibitem{flow_atlas} ATLAS Collaboration, ATLAS-CONF-2023-007
\bibitem{flow_cms}CMS Collaboration, JHEP 07 (2023) 139
\bibitem{flow_atlas_pp_jets}ATLAS Collaboration, Phys. Rev. Lett 131 (2023) 162301
\bibitem{aleph} ALEPH Collaboration, PRL 123 (2019) 212002
\bibitem{ALICE_ridge} ALICE Collaboration, arXiv:2311.14357 
\bibitem{ATLAS_flow_smallSyst} ATLAS Collaboration, Phys. Rev. C 96 (2017) 024908
\bibitem{ALICE_flow_allSys} ALICE Collaboration, Phys. Rev. Lett. 123, 142301 (2019)

\end{thebibliography}
\end{document}